# Herschel Detects a Massive Dust Reservoir in Supernova 1987A

M. Matsuura[1,2], E. Dwek[3], M. Meixner[4], M. Otsuka[4], B. Babler[5], M.J. Barlow[1], J. Roman-Duval[4], C. Engelbracht[6], K. Sandstrom[7], M. Lakićević[8,9], J.Th. van Loon[8], G. Sonneborn[3], G.C. Clayton[10], K.S. Long[4], P. Lundqvist[11], T. Nozawa[12], K.D. Gordon[4], S. Hony[13], P. Panuzzo[13], K. Okumura[13], K.A. Misselt[6], E. Montiel[6], M. Sauvage[13]

[1] Department of Physics and Astronomy, University College London, Gower Street, London WC1E 6BT, UK
[2] MSSL, University College London, Holmbury St. Mary, Dorking, Surrey RH5 6NT, UK
[3] Observational Cosmology Laboratory, Code 665, NASA Goddard Space Flight Center, Greenbelt, MD 20771, USA
[4] Space Telescope Science Institute, 3700 San Martin Drive, Baltimore, MD 21218, USA
[5] Department of Astronomy, 475 North Charter St., University of Wisconsin, Madison, WI 53706, USA
[6] Steward Observatory, University of Arizona, Tucson, AZ 85721, USA
[7] Max Planck Institut für Astronomie, Königstuhl 17, D-69117 Heidelberg, Germany
[8] Astrophysics Group, Lennard-Jones Laboratories, Keele University, ST5 5BG, UK
[9] European Southern Observatory, Karl Schwarzschild Straße 2, D-85748 Garching bei München, Germany
[10] Louisiana State University, Department of Physics & Astronomy, 233-A Nicholson Hall, Tower Dr., Baton Rouge, LA 70803-4001, USA
[11] Department of Astronomy, The Oskar Klein Centre, AlbaNova University Center, Stockholm University, SE-106 91 Stockholm, Sweden
[12] Institute for the Physics and Mathematics of the Universe, University of Tokyo, Kashiwa, Chiba 277-8583, Japan
[13] CEA, Laboratoire AIM, Irfu/SAp, Orme des Merisiers, F-91191 Gif-sur-Yvette, France

[*] To whom correspondence should be addressed; E-mail: mikako@star.ucl.ac.uk.

**We report far-infrared and submillimeter observations of Supernova 1987A, the star that exploded on February 23, 1987 in the Large Magellanic Cloud, a galaxy located 160,000 lightyears away. The observations reveal the presence of a population of cold dust grains radiating with a temperature of ~17–23 K at a rate of about 220 $L_\odot$. The intensity and spectral energy distribution of the emission suggests a dust mass of ~0.4–0.7 $M_\odot$. The radiation must originate from the SN ejecta and requires the efficient precipitation of all refractory material into dust. Our observations imply that supernovae can produce the large dust masses detected in young galaxies at very high redshifts.**

Supernovae produce most of the heavy elements found in the Universe and disperse them into their surrounding galactic environment. They chemically enrich the material from which new generations of stars and planets are formed. SN 1987A exploded on 23 February 1987 in the Large Magellanic Cloud (LMC), only 50 kpc away. Because of its proximity, it has been possible to witness its evolution from explosion to remnant. SN 1987A has thus become one of the most extensively studied extragalactic objects, with ground, airborne, and space observatories covering a wide range of the electromagnetic spectrum. Here we report far-infrared and submillimeter (submm) observations of SN 1987A.

## Previous Observations of Supernova 1987A

The detection of neutrinos from SN 1987A confirmed that the event marked the explosive death of a massive star (*1, 2*). The intensity and evolution of its UV and optical light curves showed that the SN's luminosity was powered by the radioactive decays of $^{56}$Ni and $^{56}$Co, yielding 0.06 $M_\odot$ of $^{56}$Fe (*3*). The γ-rays and X-rays from radioactive nuclei emerged only four months after the explosion (*4, 5*), suggesting that the ejecta were clumpy (*6*). The large width of near- and mid-infrared spectral lines of heavy elements (*7*) indicated that the newly synthesized material from the stellar core had undergone substantial mixing into the outer ejecta of the stellar explosion (*6*). SN 1987A was classified as a type II event on the basis of the detection of hydrogen in its optical spectrum (*8*).

Examination of plates of the region obtained before the SN explosion, allowed the detection of its progenitor, a blue supergiant (Sk–69 202), that was believed to have had an initial mass of 18–20 M$_\odot$ (*8*). The progenitor underwent extensive mass loss during its post main-sequence evolution (*8*). The detection and evolution of UV lines from highly ionized gas revealed previously ejected material that was flash ionized by the X-ray and UV radiation released when the outwardly expanding shock generated by the collapse of the stellar core broke out through the surface of the progenitor star (*9*). The presence of progenitor material was later confirmed by Hubble Space Telescope observations, which revealed a thin elliptical ring (1.6"×1.2") (*10*) containing many knots within it (*11*). The equatorial ring is part of a system of three rings forming an hour-glass like configuration centered on the explosion. The mass of the ring is estimated to be $>\sim 6 \times 10^{-2}$ M$_\odot$ (*12*).

Mid-infrared spectral and photometric observations of SN 1987A obtained with the Kuiper Airborne Observatory (*13*) revealed the presence of newly formed dust in the SN, providing direct evidence for the formation of grains in the cooling ejecta of a supernova. The dust condensation process started around day 450–600 after the explosion (*14,15*); it is thought that dust have condensed in the clumps within the ejecta (*16*). The dust mass was estimated to be at least $\sim 10^{-4}$ M$_\odot$ (*17*).

Several years after the explosion, SN 1987A had evolved from a SN remnant radiating as a result of radioactive decays, mainly $^{56}$Co and $^{44}$Ti (*3*), into a remnant where the interaction of the SN blast wave with the circumstellar medium is causing the ring to brighten up. This interaction, dominated by the collision between the blast wave and the equatorial ring, manifested through the appearance of UV-optical hot spots strung like beads around the ring (*18*). Similar ring structures were also observed at X-ray wavelengths and in the mid-infrared (*19*).

Subsequent spectral observations with the Spitzer Space Telescope on days 6200–8000 (*20*) revealed strong infrared (6–30 μm) emission arising from about $10^{-6}$ M$_\odot$ of silicate dust in the inner ring that had been collisionally heated to a temperature of 180 K by the shocked X-ray emitting gas. The dust in the ring was ejected during the pre-SN evolution of the progenitor star.

## Observations and Results

Using the Herschel Space Observatory's (*21*) imaging instruments, PACS (*22*), and SPIRE (*23*), we observed SN 1987A as part of the HERschel Inventory of The Agents of Galaxy Evolution in the Magellanic Clouds open time key program, HERITAGE (*24*). The observing survey covered 5 bands, PACS 100 and 160 μm, and SPIRE 250, 350, and 500 μm bands with angular resolutions of 6.69"×6.89", 10.65"×12.13", 18.2", 24.9" and 36.3", respectively. SN 1987A was observed on April 30[th] and August 5[th], 2010 (days 8467 and 8564 after the explosion).

The PACS and SPIRE images of the combined two epochs (Fig. 1) (*25*) reveal SN 1987A as a faint but detectable point source coincident with the VLBI measured coordinates of SN 1987A, RA: 5h 35m 27.990s, Dec: −69d 16' 11.110" (J2000) (*26*). Spatial comparison of these images with prior Spitzer detections with MIPS at 24 μm and IRAC at 8 μm (*20*) confirms the location of SN 1987A with respect to nearby interstellar features. In the PACS 100 and 160 μm images, SN 1987A is clearly detected as a point source that is well separated from the nearby interstellar medium (ISM) dust. In the SPIRE bands, the 250-μm image shows SN 1987A as a point source, but the SN begins to merge with the nearby ISM dust emission at 350 μm, and blends completely at 500 μm owing to the increasing beam size at longer wavelengths. The Hubble and Spitzer Space Telescopes and Herschel Space Observatory images (Fig. 1a, c–h) show that SN 1987A is located in a region with relatively low amounts of ISM gas and dust.

We obtained photometry from the processed Herschel images (Table 1) (*27*), assuming SN1987A is unresolved; the fluxes are plotted in Fig. 2a. There are two peaks in the spectral energy distribution (SED), one near 20 μm and the other at 150–200 μm (Fig. 2a). The total luminosity from 100–500 μm is approximately 220 L$_\odot$. The mid-infrared (5–40 μm) luminosity is 1200 L$_\odot$ (epoch of 7983 days) (*20*) and the X-ray luminosity is ~500 L$_\odot$ (8012 days) (*28*). Thus, the far-infrared/submm emission we detected is an important component of this SN's SED.

## The Origin of the far-infrared/submm Emission

The smooth blackbody-like shape of the far-infrared/submm SED (Fig. 2a) suggests continuous emission from dust. However, synchrotron emission and line emission from the same gas that produces the UV, optical and near-infrared lines are also possible.

Line emission can arise from the supernova ejecta and the circumstellar rings. The rings were first flash-ionized by the supernova, and have since recombined and cooled (*12*). To estimate the contribution from ionized lines from the inner ring, we extended the photoionization code of (*12*) to include far-infrared/submm lines. The expected line intensities of the strongest ionized lines are $0.5 \times 10^{-15}$, $1.2 \times 10^{-15}$, $0.1 \times 10^{-15}$, and $0.2 \times 10^{-15}$ erg s$^{-1}$ cm$^{-2}$ for the 88 μm [OIII], 122 μm [NII], 158 μm [CII], and 205 μm [NII] lines, respectively. These line fluxes are equivalent to a 0.2 % contribution to the total Herschel in-band fluxes; hence, photoionised lines make a negligible contribution to the far-infrared/submm fluxes. Another possible contribution to the line emission may come from the radiatively cooling gas shocked by the ejecta/ring interaction (*11*). However, it is unlikely that this contribution could be orders of magnitude higher than those from photoionised gas, because the forbidden lines are collisionally de-excited in the high-density shocked gas.

Continuum synchrotron radiation can also contribute to the emission in the Herschel bands. The synchrotron radiation has been measured at radio frequencies up to day 8014 after the explosion (*26, 29*; Fig. 2a); we extrapolated the fluxes to far-IR wavelengths using a power-law. The predicted Herschel flux densities from synchrotron radiation are two orders of magnitude lower than those we measured (Fig. 2a), implying that the emission observed by Herschel is mostly due to the continuum emission from dust.

## Dust mass

Using the flux densities and their uncertainties, we calculated the dust temperature and mass needed to account for the observations (Table 2, and Fig 2b) (*30*).

We explored four possible origins for the dust: progenitor dust, ambient ISM dust swept up by SN shocks, a light echo from ISM dust, and SN ejecta dust.

It is thought that the progenitor of SN 1987A ejected 8 M☉ of gas during its red supergiant (RS) phase prior to the SN explosion (*3*). The dust ejected this way is expected to be composed mostly of silicates because RS produce oxygen-rich dust. Adopting a gas-to-dust mass ratio of 300 (*31*) implies a silicate dust mass from the RS of 0.03 M☉. The silicate dust mass required to fit the observed emission is much higher (> 2 M☉; Table 2), equivalent to > 600 M☉ of gas. A similar evaluation rules out the possibility that dust could have been formed during another evolutionary phase of the progenitor, e.g. a luminous blue variable. It is therefore unlikely that the observed far-infrared/submm emission could have originated from dust formed by the SN progenitor.

As it expands, the SN sweeps-up dust from the ambient ISM. The SN shock speed is up to 6000 km s$^{-1}$ (*32*), and the distance already travelled by the shocks is up to 30,000 AU (0.6 arcsec). We assumed a gas density of 1 hydrogen atom per cm$^3$ in the LMC ISM and adopted a gas-to-dust mass ratio of 300. Within this volume, only a small mass of ISM dust (3 × 10$^{-6}$ M☉) will have been swept-up. This estimate is consistent with Herschel measurements of the ISM dust emission of < 10$^{-4}$ M☉, based on the sky background level in the 160-μm band (70.4±9.6 MJy sr$^{-1}$) measured for a 1.3x0.4 arcsec$^2$ region centered at the position of SN, after removal of a point spread function (PSF). Our measured dust emission requires a much higher dust mass; hence favoring emission from SN dust rather than from swept-up ISM dust.

The original SN energy pulse could potentially have been absorbed and re-emitted by interstellar dust grains to produce infrared echoes. To account for the Herschel fluxes, an infrared echo would have to originate from a dust cloud located at about 11 light years directly behind the SN, which is improbable. Moreover, this dust would have been heated by the optical and UV light, which had a luminosity of $L_{uvo}$=10$^{41}$ erg sec$^{-1}$ (*17*), implying a flux incident on the dust $F=L_{uvo}/(4\pi R^2)$, where $R$=11 light years. Taking this energy input as the dust heating rate, equilibrium with the dust cooling rate yields the dust temperatures ranging from 40 K for silicate, and 70–110 K for graphite, for a range of grain sizes from 0.01–0.1 μm (*33*). These temperatures are substantially higher than those measured by Herschel (Table 2), ruling out an echo as an explanation for the Herschel SED.

Thus, we conclude that the emission observed by Herschel is associated with cold dust in the SN ejecta.

The dust that is emitting at Herschel wavelengths could be heated by one or more of the following mechanisms: (a) by the X-ray emission arising from the interaction of the blast wave with the equatorial ring, which has a luminosity of about 500 L☉ (*20, 28*); (b) by the energy released from the long-lived radioactive $^{44}$Ti isotope present in the ejecta - our measured far-infrared/submm luminosity of 220 L☉ is about half of the predicted $^{44}$Ti luminosity at this epoch (~400 L☉) (*35*); (c) the ambient diffuse interstellar radiation field, given that the temperature of the far-infrared/submm emitting dust is very similar to that of ISM dust.

## The Origin of the Cold Dust in the Ejecta

The masses predicted for the major refractory elements (Table 3) in the ejecta of SN1987A (*3, 34*) restrict the maximum mass of dust that could have formed in the ejecta if all these elements precipitated from the gas phase (Table 4). The total available dust mass is about 0.71 M☉ in either nucleosynthesis model. The maximum mass of silicate dust of 0.51 or 0.40 M☉, depending on the models, is smaller than the 2.4 M☉ required to fit the observed far-infrared/submm flux with pure silicate emission. The same holds for carbon and iron dust. It is clear that any attempt to fit the observed spectrum with a single dust species requires a larger mass of that particular dust species than is available in the ejecta.

The far-infrared/submm fluxes could, however, be the sum of the emission from several different dust species in the ejecta. Figures 2c and 2d show the predicted far-IR SEDs arising from dust in the ejecta of SN1987A using ejecta abundances from models1 and 2, respectively (Table 4). Figure 2c and d suggest that a significant fraction of the major refractory elements are depleted onto dust, and that the total dust mass is about 0.4–0.7 M☉.

Support for the presence of large amounts of dust in the ejecta may come from HST observations of the ejecta H*a* profile in 2004 and 2010 (*36*) which show the red-ward part of the profile to be largely missing, which could be interpreted as implying that emission from material at the far side of the ejecta suffers more extinction by intervening dust than material at the front side.

The derived dust mass is about $10^3$ times larger and about twenty times colder than that measured at mid-infrared wavelengths around 600 days after the explosion, soon after it first condensed out of the ejecta (*15,16*). Far-infrared and submm observatories with Herschel's sensitivity did not exist at that time. It is possible this submm dust mass existed at day 600. However, it is also possible that this dust mass grew by accretion over the past 20 years, or that initially optical/infrared-thick dust-bearing clumps may have become optically thin at far-infrared and submm wavelengths. Overall, the observations provide evidence for the efficient formation and growth of massive quantities of dust in SN ejecta.

Our result could impact on the understanding of the evolution of dust in galaxies. In the LMC, the rate of type II SNe is estimated to be one per 300 years (*37*). If all type II SNe form typically 0.4 M☉ of dust, the overall SN dust injection rate into the ISM would be more than $10^{-3}$ M☉ year$^{-1}$. This is about ten times higher than the overall dust inputs measured from asymptotic giant branch stars in the LMC (*38, 39*). With such a high efficiency and with a dust lifetime of 2–4×$10^8$ years (*40*), it would be possible to explain a substantial fraction of the dust in the LMC (about $1 \times 10^6$ M☉; *41, also 31*) by a stellar origin.

The large amount of dust inferred to be present in the ejecta of SN 1987A is consistent with that required to explain the dust masses in high-redshift galaxies (*42, 43*), i.e. 0.1–1.0 M☉ of dust per SN, if the dust is not significantly destroyed during its injection into the ISM and subsequent encounters with interstellar shocks. Supernovae may therefore be significant contributors of dust detected in such galaxies.

**Acknowledgements:** Herschel is an ESA space observatory with science instruments provided by European-led Principal Investigator consortia and with important participation from NASA. We acknowledge financial support from the NASA Herschel Science Centre, JPL contracts no. 1381522 and 1381650. We acknowledge the contributions and support of the European Space Agency (ESA), the PACS and SPIRE teams, the Herschel Science Centre, the NASA Herschel Science Centre and the PACS and SPIRE instrument control centres, without which none of this work would have been possible. We also thank the HERITAGE team for various inputs on the data reduction. M.M. appreciates many discussions and inputs from Drs. Hirashita, Sakon, and Wesson and Profs. Nomoto, and Swinyard. M.M. acknowledges the award of a UCL Institute of Origins Fellowship and M.L. acknowledges an ESO/Keele studentship. We thank to Dr. Panuzzo for making a nice image for us. We thank the Lorentz Center for their support in finishing the paper.

| Band | Flux (mJy) |
|---|---|
| PACS 100 μm | 70.5 ± 8.5 |
| PACS 160 μm | 125.3 ± 16.1 |
| SPIRE 250 μm | 131.7 ± 12.1 |
| SPIRE 350 μm | 49.3 ± 6.5 |
| SPIRE 500 μm | <57.3* |

Table 1: The measured PACS and SPIRE flux densities of SN 1987A. We added in quadrature the uncertainties from the source extraction and the absolute errors in the flux calibration, which are estimated to be 10% for the PACS bands and 7% for the SPIRE bands (*47, 48*). * 3-sigma upper limit.

| Dust Species | $M_d$ (M$_\odot$) | $T_d$ (K) |
|---|---|---|
| Amorphous carbon | 0.35 ± 0.06 | 21.2 ± 0.7 |
| Silicate | 2.4 ± 0.4 | 17.7 ± 0.5 |
| Iron ($a$=0.1 μm) | 3.4 ± 0.6 | 19.2 ± 0.7 |
| Iron ($a$=0.5 μm) | 0.34 ± 0.06 | 25.7 ± 0.9 |

Table 2: Dust temperatures ($T_d$) and corresponding dust masses ($M_d$) derived by fitting the whole of the far-infrared/submm emission with a single dust species (*30*). Quantities are insensitive to grain radius, except metallic iron. The absorption efficiency of metallic iron depends on grain radius ($a$), and the table represents the results for grain radii of 0.1 and 0.5 μm.

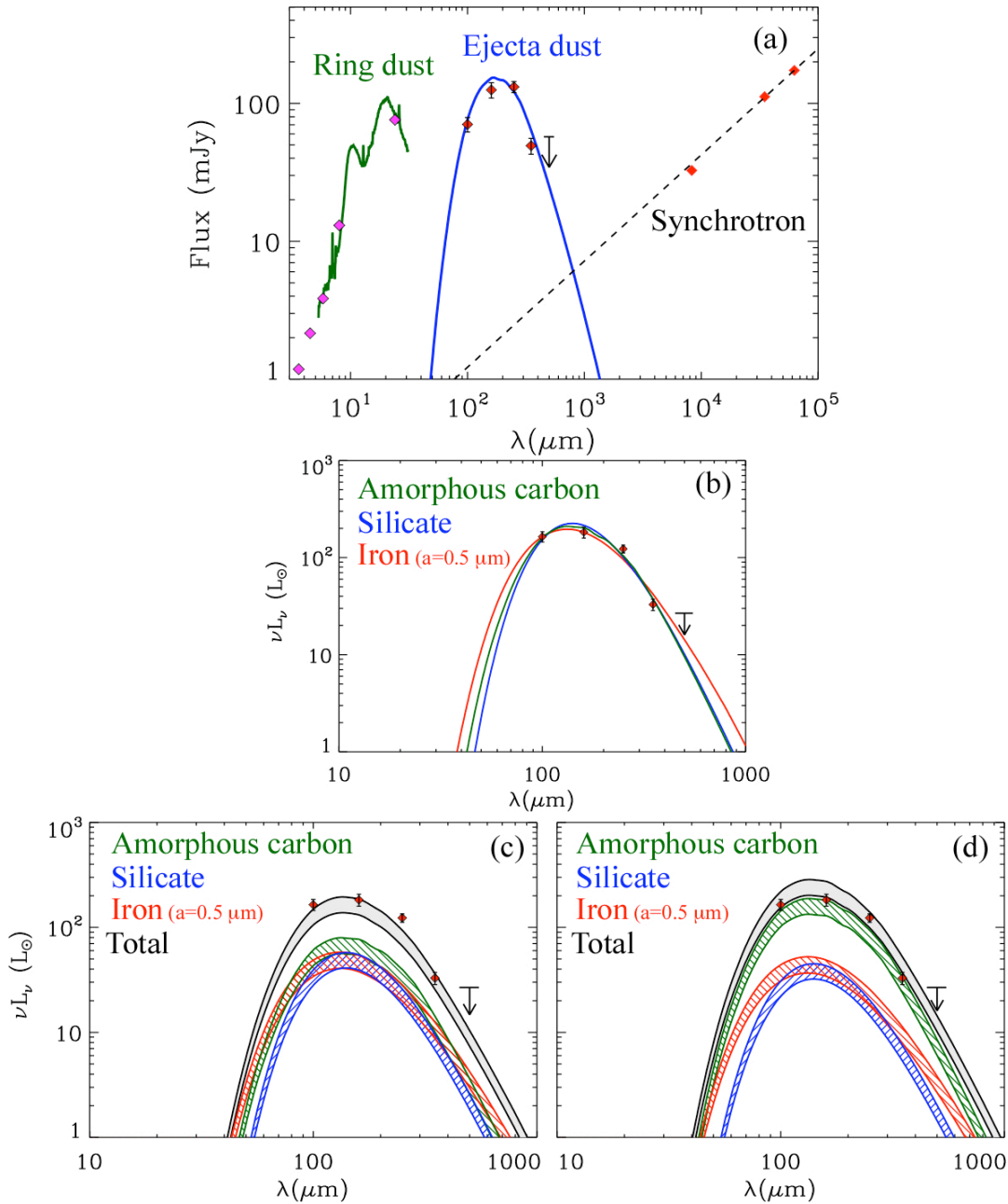

Figure 1: The Herschel images of SN 1987A together with the *Spitzer* infrared *(20)* and the *Hubble* optical *(56, 57)* images. North is top and east is left. The two vertical white lines indicate the position of SN 1987A measured from radio observations. (**Inset i**) Background-subtracted 350-μm image, where the background is estimated from the 250-μm residual image after the subtraction of the point spread function at the position of SN 1987A. The PSFs shows the resolution of the Herschel instruments. Panel (b) shows an enlarged HST optical image, indicating the morphology of the SN remnant. [Source: panel (a), the Hubble Heritage Team (AURA/STScI/NASA); panel (b), NASA, ESA, P. Challis and R. Kirshner (Harvard-Smithsonian Center for Astrophysics)].

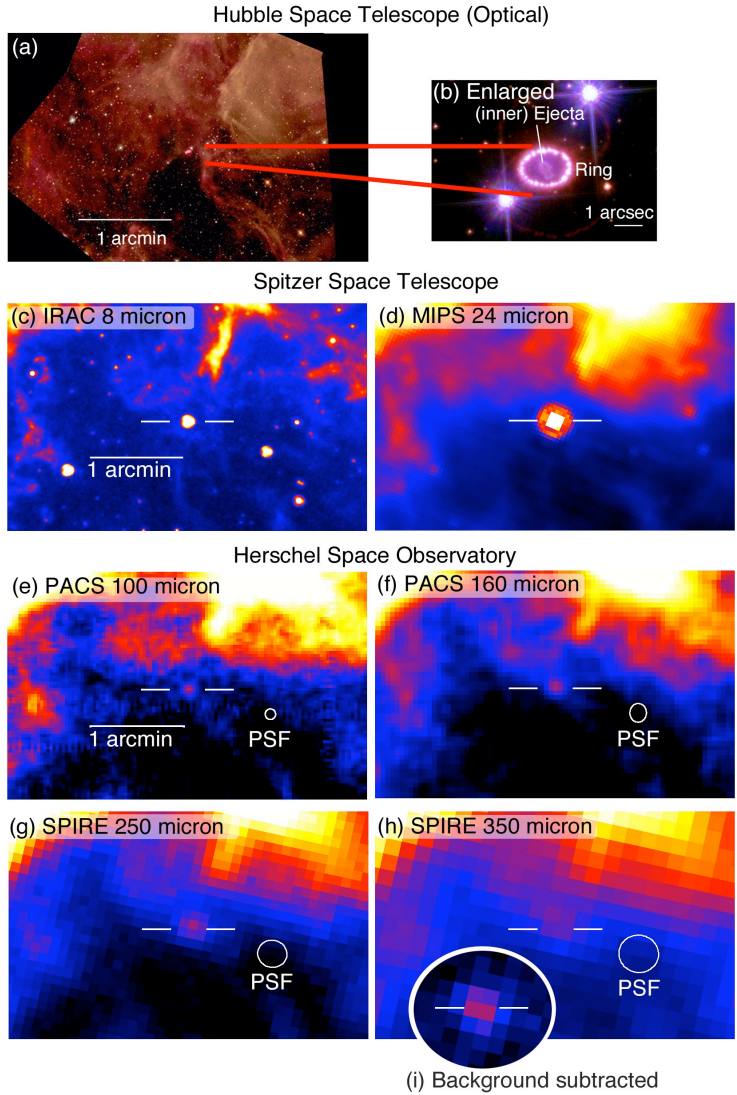

Figure 2: (a) The infrared spectral energy distribution of SN 1987A. Herschel detected SN 1987A from 100 to 350 μm. An upper limit is given at 500 μm. The other observational data were collected from the literature: 5.8–24 μm photometric data and 5–30 μm spectra (*20*), which measures the warm dust emission; and the radio continuum (*26, 29*), which traces the synchrotron emission. (b): model fits to the far-infrared/submm dust emission using different single dust species. Parameters are given in Table 2. (c and d): The sum of the contributions from each dust species to the far-infrared/submm spectrum of SN 1987A. The flux at each wavelength is that derived from fitting the observed spectrum with each dust species alone (Table 2), scaled by the fraction of its mass available in the ejecta from abundance considerations (Table 4). The model 1 and model 2 abundances in Table 4 correspond to (c) and (d), respectively. The hatched areas represent the range of possible fluxes from each species, with the lower and upper boundaries corresponding, respectively, to the scaling of the minimum and maximum and dust masses (uncertainties in Table 2) required to produce the observed spectrum.

| Elements | Mass ($M_\odot$) | |
|---|---|---|
| | Model 1 | Model 2 |
| C | 0.114 | 0.26 |
| O | 1.48 | 0.24 |
| Mg | 0.183 | 0.04 |
| Si | 0.101 | 0.14 |
| Fe | 0.083 | 0.075 |

Table 3: Model prediction of the mass of representative refractory elements synthesized in the SN 1987A ejecta. Isotopes which contribute a large fraction of the total atomic mass were accounted for. Model 1: theoretical calculations of elemental abundance from (*3*). Model 2: calculations from an updated version of (*34*).

| Dust Species | $m_d$ ($M_\odot$) | |
|---|---|---|
| | Model 1 | Model 2 |
| Amorphous carbon | 0.11 | 0.26 |
| Silicate | 0.52 | 0.37 |
| Iron | 0.08 | 0.08 |
| Total | 0.71 | 0.71 |

Table 4: The dust mass assumes 100% dust condensation of the available elemental mass ($m_d$). The range of dust masses reflects the difference in compositions in Table 3. All silicates are assumed to be in the form of MgO and $SiO_2$ dust. The mass of carbon dust assumes that no significant fraction of carbon is locked up in CO molecules.